\documentclass[10pt,conference]{IEEEtran}
\IEEEoverridecommandlockouts
\usepackage{cite}
\usepackage{url}
\usepackage{amsmath,amssymb,amsfonts}
\usepackage{algorithmic}
\usepackage{graphicx}
\usepackage{textcomp}
\usepackage{listings}
\usepackage{hyperref}
\usepackage{xcolor}
\usepackage[shortlabels]{enumitem}
\def\BibTeX{{\rm B\kern-.05em{\sc i\kern-.025em b}\kern-.08em
    T\kern-.1667em\lower.7ex\hbox{E}\kern-.125emX}}

\begin{document}

\title{Design choices made by LLM-based test generators prevent them from finding bugs}

\author{
\IEEEauthorblockN{Noble Saji Mathews}
\IEEEauthorblockA{
University of Waterloo\\Canada \\
noblesaji.mathews@uwaterloo.ca}
\and
\IEEEauthorblockN{Meiyappan Nagappan}
\IEEEauthorblockA{
University of Waterloo\\Canada \\
mei.nagappan@uwaterloo.ca}
}

\maketitle

\begin{abstract}
There is an increasing amount of research and commercial tools for automated test case generation using Large Language Models (LLMs). 
This paper critically examines whether recent LLM-based test generation tools, such as Codium CoverAgent and CoverUp, can effectively find bugs or unintentionally validate faulty code. 
Considering bugs are only exposed by failing test cases, we explore the question: can these tools truly achieve the intended objectives of software testing when their test oracles are designed to pass?
Using real human-written buggy code as input, we evaluate these tools, showing how LLM-generated tests can fail to detect bugs and, more alarmingly, how their design can worsen the situation by validating bugs in the generated test suite and rejecting bug-revealing tests. 
These findings raise important questions about the validity of the design behind LLM-based test generation tools and their impact on software quality and test suite reliability.
\end{abstract}


\section{Introduction}


Software testing remains a critical yet time-consuming aspect of software development. With the advent of Large Language Models (LLMs), there has been a significant shift towards automating various software engineering tasks, including test case generation. Tools like GitHub Copilot, which serves millions of developers \cite{dohmke2023economic}, are increasingly being adopted for automated test generation. This trend is further visible in recent research efforts \cite{pizzorno2024coverup, el2024using, dakhel2024effective} and commercial tools like CodiumAI's CoverAgent \cite{githubGitHubQodoaiqodocover}.
While there are various approaches to automated test generation, including fuzzing and crash-based testing, our study focuses on LLM-based tools that generate complete test cases with oracles and assertions.

LLM-based test generation tools \cite{githubGitHubQodoaiqodocover, pizzorno2024coverup} promise to improve code coverage and generate tests by observing the existing implementation and filtering out or attempting to “fix” failing tests.
Such an approach raises critical concerns about their effectiveness in achieving the primary goal of software testing: detecting bugs.
This challenge aligns with the broader ``test oracle problem'', where distinguishing correct from faulty behaviour requires reliable oracles  \cite{barr2014oracle}. 

This study explores how current LLM-based test generation tools handle buggy implementations, highlighting the risks of relying on automatically generated tests.
We present empirical evidence demonstrating how these tools can inadvertently validate incorrect behaviour through their test generation and filtering mechanisms. 
Our findings suggest that current LLM-based test generation approaches may be fundamentally misaligned with the core objective of software testing (revealing bugs), pointing to a critical gap that must be addressed before these tools can be reliably integrated into software testing practices.

\section{Related Work} \label{related_work}
Automated test generation has been extensively studied, with approaches ranging from specification-based to search-based software testing (SBST) \cite{fraser2011evolutionary}. For Python specifically, Pynguin \cite{lukasczyk2022pynguin} employs SBST techniques, using genetic algorithms to evolve test cases for improved coverage. Recent work has enhanced these traditional approaches with LLMs, as demonstrated by CodaMosa \cite{lemieux2023codamosa}, which uses LLMs to overcome stalled search states in Pynguin's genetic algorithm. It however explicitly refrains from using LLMs for generating oracles and only uses them to generate inputs. 
While Pynguin itself can generate oracles by running the program on the input, this approach relies solely on runtime behaviour rather than leveraging LLMs for oracle generation.

The success of LLMs in Software Engineering has led to the development of several specialized test-generation tools. MuTAP \cite{dakhel2024effective} focuses on mutation testing, iteratively prompting LLMs to generate assertions for surviving mutants. MuTAP suffers from the same problems highlighted for other tools in this study as it follows a similar filtering mechanism. However, MuTAP is not designed to be easily executable on an arbitrary codebase hence we instead include a much more recent research tool Coverup \cite{pizzorno2024coverup}. Tools like Codium CoverAgent and CoverUp take different approaches to coverage-driven test generation, with CoverUp specifically using coverage measurements to guide LLM prompting and maintaining an iterative dialogue for test refinement. These tools represent a shift from traditional automated testing approaches to LLM-assisted generation, though their effectiveness and reliability remain areas of active research.

Recent work by Huang et al. raises critical concerns about LLM-based test generation, showing that incorrect code can significantly mislead LLMs in generating reliable and bug-revealing tests \cite{huang2024rethinking}. Our work builds upon these findings by specifically examining if and how LLM-based test generation tools might not only fail to detect bugs, but potentially validate faulty code through their design.

\section{Methodology}

\begin{figure*}[ht]
    \centering
    \includegraphics[width=\linewidth]{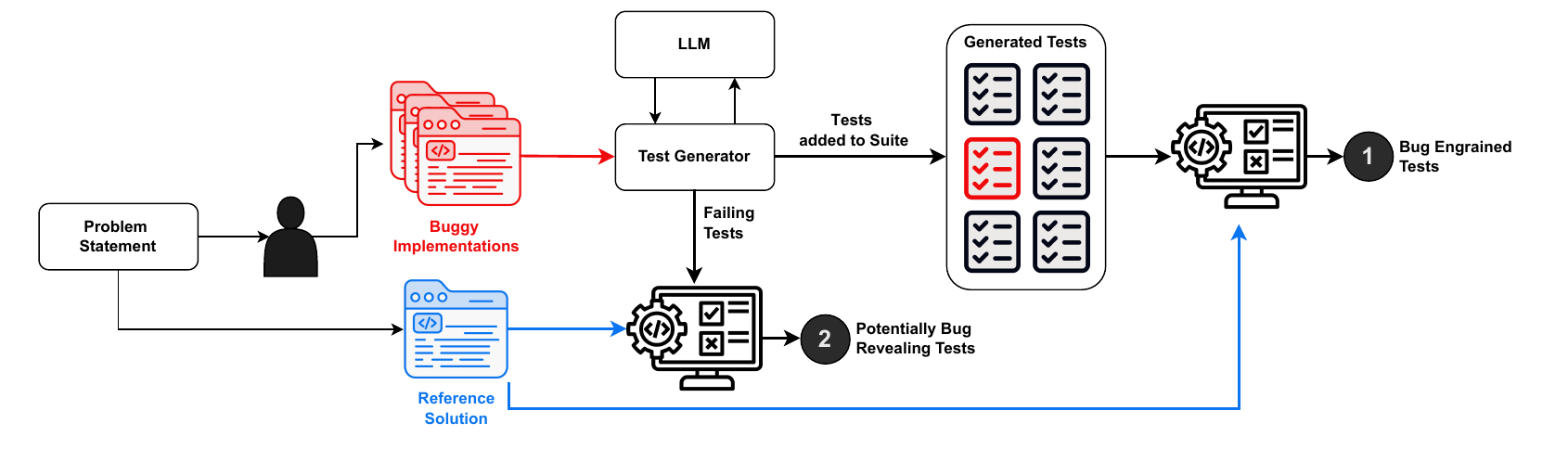}    
    \caption{Overview of the evaluation pipeline}
    \label{fig:Approach}
\end{figure*}

We design our experiments, such that we can analyze both the passing and failing tests generated by recent LLM-based test generation tools. We aim to understand how different approaches, adopted by these tools affect the generated test suite and if these tests would validate bugs or unintended behaviour when given potentially buggy code.

\subsection{Dataset} \label{dataset}

For our study, we utilize the Refactory dataset \cite{yang2019refactory}, a benchmark for Python bug repairing. This dataset is particularly suitable for our exploration as it contains real-world buggy code snippets generated by humans. We chose to focus on Python as it is widely used across various domains and as a best-case scenario as LLMs have been shown to demonstrate better performance in synthesizing Python code compared to other languages \cite{dakhel2024effective}. Refactory comprises submissions from 361 undergraduate students for 5 programming assignments in an introductory Python course at the National University of Singapore. It contains 2,442 correct and 1,783 buggy program attempts. All programs in Refactory are self-contained and mostly involve the creation of a single function, allowing test generation tools to focus on these alone, reducing factors that could influence the results we highlight in this paper.

This dataset allows us to evaluate test case generation on actual human-written buggy implementations and a reference solution based on the same requirement (problem statement here) and compare them across different tools, reflecting the reality that code is often imperfect and prone to bugs. This dataset allows us to observe and comment on the performance of these tools in a controlled yet realistic setting, without the added complexity of seeded bugs in large-scale projects.

To prepare the dataset for our experiments, we performed the following steps:    



\begin{enumerate}
    \item We retained cases with at least one passing and one failing test to focus on implementations that are executable and partially meet the requirements of the original test suite.

    \item The existing input-output tests in the dataset were converted to the format expected by the unit testing library PyTest to ensure compatibility with the tools under evaluation, which require a configured testing framework for execution.

    \item To keep our analysis tractable, we removed duplicates and selected a subset of cases where the test coverage, based only on passing tests, was less than 100\%. This also ensures that the selected cases provide opportunities for the tools to improve test coverage.
\end{enumerate}


Due to the complexity introduced by multiple target functions, we excluded assignment 2 from our analysis which requires 3 functions to be implemented, unlike the other problems. We also retained details about which tests passed and failed for each buggy implementation to aid with manual inspection of the results. After applying these filters, our final dataset consisted of 287 data samples across 4 questions, which we used to explore our research questions.

\subsection{Test Generation Tools}

For our study, we picked three prominent LLM-based test generation tools: GitHub Copilot, Codium CoverAgent, and CoverUp that represent different approaches to automated test generation using LLMs that are available to Python developers today. We go over other past approaches in Section~\ref{related_work}.

\vspace{4mm}
\subsubsection{GitHub Copilot}

GitHub Copilot is a widely adopted AI-powered code completion and generation tool that integrates with various development environments. While primarily known for code generation, Copilot also offers capabilities for generating unit tests. As of writing it operates on a pure generation model, suggesting test cases based on the context of the code in the editor or highlighted snippets. Copilot can propose input parameters, expected outputs, and assertions for functions, potentially aiding in creating test cases for edge cases and boundary conditions \cite{el2024using}.

However, it's important to note that Copilot generates tests without executing them or verifying their correctness against the implementation. While potentially time-saving, this approach may lead to the generation of tests that don't compile, fail to run, or don't increase code coverage.

\vspace{4mm}
\subsubsection{Codium CoverAgent}

Codium CoverAgent (now renamed to qodo-cover) is an open-source implementation\footnote{Qodo offers an enterprise version which we don't have access to} inspired by Meta's TestGen-LLM research \cite{alshahwan2024automated} which adopts an ``Assured LLM-based Software Engineering'' approach aiming to augment existing test suites with additional test cases while ensuring no regression. The tool employs an iterative process:

\begin{enumerate}[a)]
\item Measuring existing test suite coverage (if any)
\item Generate multiple test cases
\item Filter out tests that don't compile or run
\item Discard tests that don't pass on the current code
\item Retain only tests that increase code coverage
\item Repeat the process while disincentivizing the creation of tests like the failing ones  
\end{enumerate}

This approach attempts to address common issues in LLM-based test generation, such as hallucination and the generation of low-value or non-compiling tests. However, by design, it may inadvertently reinforce existing bugs in the code by only keeping tests that pass on the current, potentially faulty, implementation. Further, the explicit prompt to not generate tests similar to the failing cases might end up with bug-prone regions being ignored.

\vspace{4mm}
\subsubsection{CoverUp}

CoverUp represents a more recent approach in LLM-based test generation, focusing on creating high-coverage regression tests for Python \cite{pizzorno2024coverup}. It employs an Agent \cite{xi2023rise} style approach that interleaves coverage analysis with LLM interactions. The process involves:

\begin{enumerate}[a)]
\item Measuring existing test suite coverage (if any)
\item Identifying uncovered code portions
\item Prompt the LLM to generate tests for each uncovered portion
\item Verifying that new tests run and increase coverage
\item Refining prompts based on updated coverage information
\item In case of suite failure, test failure details are used to further engage in dialogue with the LLM
\item Continue onto the next uncovered portion (if any) after the max iterations are reached
\end{enumerate}

CoverUp aims to guide the LLM towards generating tests that specifically target uncovered code, potentially leading to higher overall coverage compared to other approaches. It also provides the LLM with the ability to collect context about different parts of the codebase if it wishes to do so. As compared to CoverAgent, this design is potentially more susceptible to creating bug-engrained tests as it attempts to ``repair'' failing test suites iteratively.

These tools were chosen for our study due to their distinct approaches to test generation and their potential impact on software testing practices. While each tool claims to improve the efficiency and effectiveness of test creation, our research aims to critically evaluate their performance, particularly in the context of potentially buggy implementations. By examining these tools, we seek to understand the implications of relying on LLM-generated tests for bug detection and code validation.

\subsection{Evaluation Pipeline}
Figure~\ref{fig:Approach} provides an overview of the evaluation pipeline adopted for this study. We utilize the 287 buggy samples filtered from the Refactory dataset as described in Section~\ref{dataset} and also collect the reference solutions associated with each of the four problem statements. We deliberately utilize the default parameters wherever possible and only make the necessary changes to have the tools work on our extracted samples. This also means we stick with the default LLMs adopted by the tools which at the time of writing is OpenAI's GPT4O (Version 2024-08-06) for CoverAgent and Coverup. The exact LLM used is not very important in our study as we attempt to focus on the design and objectives of pipelines within these tools rather than the power of the LLM itself, using GitHub copilot as sort of a moving baseline being a popular pure generative approach to the problem providing the bare minimum functionality for ``one-click'' test generation with machinery to collect the required context from the project. We had each test generator write tests from scratch effectively populating an empty test suite. A more detailed breakdown of the automated test generation setups we used in each case is provided below:

\subsubsection{GitHub Copilot}
    
We implemented an automated framework to evaluate GitHub Copilot's performance in generating unit tests. Using Playwright, the script programmatically interacts with GitHub Codespaces to ensure a consistent and reproducible process, this also helps us execute Copilot commands as would be used by developers within VSCode without introducing any variability. The framework begins by setting up the environment, including authentication and session persistence, to streamline task execution.

Each task, described in a dataset with function signatures and sample content, is processed iteratively one code snippet at a time, resetting the environment at the end of each iteration. Source and test files are created automatically, and the specific method to be tested is selected within the editor to provide context for Copilot. The Copilot command \texttt{'Github Copilot: Generate Tests'} is invoked to trigger test generation. User interactions, such as accepting or applying generated suggestions, are automated to accept all suggestions. Generated test files are then organized into a designated directory for further evaluation.

\subsubsection{Codium CoverAgent}

Our script iterates through directories containing buggy code samples and their corresponding test files which contain blank templates with a dummy empty test but can also be instantiated with existing tests if required. It invokes CoverAgent using the following command structure:

\begin{lstlisting}
    cover-agent --source-file-path [source] --test-file-path [test] --code-coverage-report-path [report] --test-command [pytest] --coverage-type [cobertura] --desired-coverage [100]
\end{lstlisting}

CoverAgent is thus tasked to generate tests iteratively until the desired code coverage or the maximum number of iterations is reached (defaults to 5). Each iteration filters out tests that fail to compile or run, retaining only those that pass and contribute to increased coverage. We instrument the implementation of CoverAgent such that we gain access to all tests generated and also collect additional information surrounding rejection. Logging is enabled to capture execution details and any errors encountered.

\subsubsection{CoverUp}

Our automated pipeline integrates CoverUp by processing buggy code samples and preparing a structured environment for its execution. For each code sample, the script creates a dedicated directory structure with separate \texttt{src} and \texttt{tests} folders, copying the source files and optionally generating blank test templates similar to the setup for CoverAgent.

The tool is executed using the command:

\begin{lstlisting}[language=bash]
coverup --tests-dir [tests] --package-dir [src]
\end{lstlisting}

CoverUp iteratively measures coverage, identifies uncovered code segments, and generates tests to target those segments. It attempts to cover each uncovered segment It refines the process by analyzing test failures and adjusting its prompts to improve coverage. Success and failures on test suite execution are collected for analysis, and the process continues until the tool exhausts its iterations or achieves the desired coverage.


\section{Results}
Table \ref{tab:results} shows the outcomes of evaluating test suites generated by each tool against both the buggy implementations (OG) and reference solutions (REF) in each case. For each tool, we analyze how the generated test suites behave when executed against these two implementations, capturing four possible outcomes that help us understand if the tests are effectively identifying bugs or inadvertently validating them. These are:

\begin{table*}[tbh]
\renewcommand{\arraystretch}{1.5} 
\caption{Evaluation Results of Generated Test Suites}
\label{tab:results}
\begin{tabular}{lcccc}
\hline
\textbf{Tool} & \textbf{OG Failed, REF Success} & \textbf{OG Failed, REF Failed} & \textbf{OG Success, REF Success} & \textbf{OG Success, REF Failed} \\
\hline
GitHub Copilot & 194 & 66 & 15 & 12 \\
CodiumAI & 470* & 1296* & 116 & 171 \\
CoverUp & 400* & 235* & 29 & 62 \\
\hline
\end{tabular}

\vspace{0.5em} 
\small
\textbf{OG} - Original buggy implementation used to generate the suite 
|
\textbf{REF} - Reference Implementation \\
* These test suites were rejected during the generation process
\end{table*}

\begin{itemize}
    \item \textbf{OG Failed, REF Success}: Tests that fail on the buggy implementation, but pass on the reference solution. These represent potentially valuable tests that correctly identify bugs. For GitHub Copilot, 194 test cases (67.6\% of generated tests) fall into this category. However, both CodiumAI and CoverUp reject such tests during their generation process, with 470 and 400 such tests being discarded respectively.
    
    \item \textbf{OG Failed, REF Failed}: Tests that fail on both implementations. These typically represent poorly generated tests that either don't compile or contain invalid assertions. GitHub Copilot generated 66 such cases (23\% of its output), while CodiumAI and CoverUp rejected 1296 and 235 such tests respectively during their filtering process.
    
    \item \textbf{OG Success, REF Success}: Tests that pass on both implementations. These tests exercise correct behaviour common to both implementations. This category represents the smallest portion for all tools: 15 cases (5.2\%) for Copilot, 116 cases (40.4\%) for CodiumAI, and 29 cases (31.9\%) for CoverUp's suites.
    
    \item \textbf{OG Success, REF Failed}: The most concerning category, is tests that pass on buggy code but fail on correct implementations. These tests effectively validate bugs by treating incorrect behaviour as expected. We found 12 such cases (4.2\%) in Copilot's output, while CodiumAI and CoverUp produced 171 (59.6\%) and 62 (68.1\%) such cases respectively in their final test suites.
\end{itemize}

It should be noted when looking at these numbers that in the first 2 cases (with OG Failed), CoverUp and CoverAgent do not generate test suites and reject all tests. We collect these ``failed tests'' by tapping into the tool's inner workings. Additionally, CoverAgent disincentivizes the creation of tests similar to the failed cases.
CoverUp failed to generate a test suite for 196 cases (68.3\% of the dataset) as its repair attempts were unsuccessful. This high failure rate, combined with the concerning proportion of tests that validate bugs in the final suites (68.1\%), suggests that current approaches to test generation might be fundamentally flawed in their ability to handle buggy implementations and detect bugs effectively.

\section{Discussion}

The results demonstrate a concerning trend: tools that implement filtering mechanisms (CodiumAI and CoverUp) tend to produce final test suites that are more likely to validate bugs compared to pure generation approaches like GitHub Copilot. This is evidenced by the higher percentage of ``OG Success, REF Failed'' cases in their final outputs, suggesting that their filtering and fixing mechanisms might be counterproductive on real code that tends to have bugs.

Simple experiments can expose fundamental issues with current LLM-based test generation approaches. Consider the example below:
\begin{lstlisting}[language=Python]
# Buggy Implementation
def find_sum(a,b):
    return (a+b+1)  # Bug: Adds 1 to sum

# Copilot Generated Test Suite
def test_find_sum(self):
    self.assertEqual(find_sum(1, 2), 4)
    self.assertEqual(find_sum(-1, 1), 1)
    self.assertEqual(find_sum(0, 0), 1)
    self.assertEqual(find_sum(100, 200), 301)
    self.assertEqual(find_sum(-5, -5), -9)
\end{lstlisting}

In this case, the function incorrectly adds 1 to every sum, yet the generated test suite validates this incorrect behaviour by asserting the buggy outputs as expected results. While GitHub Copilot might generate such tests through pure generation, tools like CodiumAI and CoverUp systematically retain these bug-validating tests while rejecting potentially valuable tests that could expose the bug. 
While one might argue that the incorrect behaviour could align with some specific use case, the lack of a docstring or other semantic cues leaves the intended functionality ambiguous. Unlike Copilot, which can be directed to write specific tests, tools like CodiumAI and CoverUp infer requirements from the code itself and make their sole objective to increase coverage. This reliance on inferred intent introduces challenges when bugs are ingrained in the code. Such bugs may distort how developers interpret test failures, as it becomes unclear whether the issue lies in a faulty generated test (requiring updates to both the test and the code) or in the code itself (necessitating a code fix and a new test). This diverges from the typical development process, where test failures primarily indicate code issues.

To ascertain the implications of this behaviour in the real world, we manually ran and inspected the output of these tools on a few select issues from SWEBench \cite{jimenez2024swebench}, a dataset used to benchmark LLMs on their ability to solve Software Issues. For instance, we analyzed a bug in SymPy's crypto module (\href{https://github.com/sympy/sympy/issues/16884}{Issue \#16884}) where an incorrect Morse code mapping for the digit ``1'' persisted for over three years (January 2016 to May 2019). The bug involved mapping ``----'' to ``1'' instead of the correct ``.----''. When presented with this buggy implementation, both CoverUp and CoverAgent discarded tests involving the encoding of ``1'', effectively masking the bug. Highlighting how the current approach of filtering ``failing'' tests can systematically hide important edge cases and validation scenarios.

Another issue we observed was with the use of code coverage as the primary objective for test generation. Consider \href{https://code.djangoproject.com/ticket/34243}{Django issue \#34243}, where the \texttt{timesince()} function crashed with timezone-aware dates due to incorrect timezone handling. While the LLM-based tools generated test suites achieving high coverage, they failed to test crucial functionality across different timezones as test generation is prematurely stopped once coverage targets were met. We also note that when generating from scratch the generated suite was much less comprehensive compared to the human-written test class. We thus strongly believe that as we move closer toward AI-native software engineering, code should be written from requirements which may be expressed through tests in line with Test Driven Development \cite{mathews2024test}. The design of test generation tools thus probably needs to be such that they assist the developer in writing quality tests rather than attempting to infer the requirements of the user.

\section{Threats to Validity}
The primary threat to our study's validity stems from the dataset composition. While using student-written buggy implementations from Refactory provides us with controlled examples where we have both incorrect implementations and verified reference solutions, these may not fully represent the complexity of bugs found in production systems. However, our findings from real-world examples suggest that the identified issues persist in production environments. Another consideration is our implementation of automated test execution pipelines for each tool. To mitigate this, we maintained default configurations wherever possible and documented necessary modifications in our replication package. While the inherent non-deterministic nature of LLM outputs could affect individual results, our focus on systematic issues in test generation pipelines rather than LLM capabilities means this variability does not significantly impact our main findings.

Our evaluation assumes that the reference solutions represent correct implementations, and our interpretation of test failures and successes might be affected by implementation-specific details of each tool's test generation pipeline. To address these concerns, we manually verified a sample of the reference solutions and focused our analysis on clear-cut cases where tests either definitively validate bugs or fail to detect them, rather than analyzing borderline cases that could be subject to interpretation. 

\section{Conclusion}

Our study reveals fundamental issues in the design of current LLM-based test generation tools that could significantly impact software quality. Through empirical evaluation of three tools GitHub Copilot, CodiumAI CoverAgent, and CoverUp, we demonstrate how certain design choices can lead to the validation of bugs rather than their detection, with up to 68.1\% of their final test suites validating bugs by passing on incorrect implementations while failing on correct ones. The use of code coverage as a primary objective and the systematic filtering of failing tests
can mask critical bugs and create a false sense of security.
While these tools show promise in reducing the effort required for test creation, their current design philosophies need significant revision to make use of LLMs for oracle generation and better serve the fundamental purpose of software testing. We strongly believe that code should be written based on requirements and not the other way around. Until these issues are addressed, developers should exercise caution when relying on automatically generated tests, particularly during active development phases when code is most likely to contain bugs.




\bibliographystyle{IEEEtran}
\bibliography{main}

\begin{thebibliography}{10}
\providecommand{\url}[1]{#1}
\csname url@samestyle\endcsname
\providecommand{\newblock}{\relax}
\providecommand{\bibinfo}[2]{#2}
\providecommand{\BIBentrySTDinterwordspacing}{\spaceskip=0pt\relax}
\providecommand{\BIBentryALTinterwordstretchfactor}{4}
\providecommand{\BIBentryALTinterwordspacing}{\spaceskip=\fontdimen2\font plus
\BIBentryALTinterwordstretchfactor\fontdimen3\font minus \fontdimen4\font\relax}
\providecommand{\BIBforeignlanguage}[2]{{%
\expandafter\ifx\csname l@#1\endcsname\relax
\typeout{** WARNING: IEEEtran.bst: No hyphenation pattern has been}%
\typeout{** loaded for the language `#1'. Using the pattern for}%
\typeout{** the default language instead.}%
\else
\language=\csname l@#1\endcsname
\fi
#2}}
\providecommand{\BIBdecl}{\relax}
\BIBdecl

\bibitem{dohmke2023economic}
T.~Dohmke, ``The economic impact of the ai-powered developer lifecycle and lessons from github copilot,'' 2023.

\bibitem{pizzorno2024coverup}
J.~A. Pizzorno and E.~D. Berger, ``Coverup: Coverage-guided llm-based test generation,'' \emph{arXiv preprint arXiv:2403.16218}, 2024.

\bibitem{el2024using}
K.~El~Haji, C.~Brandt, and A.~Zaidman, ``Using github copilot for test generation in python: An empirical study,'' in \emph{Proceedings of the 5th ACM/IEEE International Conference on Automation of Software Test (AST 2024)}, 2024, pp. 45--55.

\bibitem{dakhel2024effective}
A.~M. Dakhel, A.~Nikanjam, V.~Majdinasab, F.~Khomh, and M.~C. Desmarais, ``Effective test generation using pre-trained large language models and mutation testing,'' \emph{Information and Software Technology}, vol. 171, p. 107468, 2024.

\bibitem{githubGitHubQodoaiqodocover}
``{G}it{H}ub - qodo-ai/qodo-cover: {Q}odo-{C}over: {A}n {A}{I}-{P}owered {T}ool for {A}utomated {T}est {G}eneration and {C}ode {C}overage {E}nhancement! --- github.com,'' \url{https://github.com/qodo-ai/qodo-cover}, [Accessed 13-12-2024].

\bibitem{barr2014oracle}
E.~T. Barr, M.~Harman, P.~McMinn, M.~Shahbaz, and S.~Yoo, ``The oracle problem in software testing: A survey,'' \emph{IEEE transactions on software engineering}, vol.~41, no.~5, pp. 507--525, 2014.

\bibitem{fraser2011evolutionary}
G.~Fraser and A.~Arcuri, ``Evolutionary generation of whole test suites,'' in \emph{2011 11th International Conference on Quality Software}.\hskip 1em plus 0.5em minus 0.4em\relax IEEE, 2011, pp. 31--40.

\bibitem{lukasczyk2022pynguin}
S.~Lukasczyk and G.~Fraser, ``Pynguin: Automated unit test generation for python,'' in \emph{Proceedings of the ACM/IEEE 44th International Conference on Software Engineering: Companion Proceedings}, 2022, pp. 168--172.

\bibitem{lemieux2023codamosa}
C.~Lemieux, J.~P. Inala, S.~K. Lahiri, and S.~Sen, ``Codamosa: Escaping coverage plateaus in test generation with pre-trained large language models,'' in \emph{2023 IEEE/ACM 45th International Conference on Software Engineering (ICSE)}.\hskip 1em plus 0.5em minus 0.4em\relax IEEE, 2023, pp. 919--931.

\bibitem{huang2024rethinking}
D.~Huang, J.~M. Zhang, M.~Du, M.~Harman, and H.~Cui, ``Rethinking the influence of source code on test case generation,'' \emph{arXiv preprint arXiv:2409.09464}, 2024.

\bibitem{yang2019refactory}
Y.~Hu, U.~Z. Ahmed, S.~Mechtaev, B.~Leong, and A.~Roychoudhury, ``Re-factoring based program repair applied to programming assignments,'' in \emph{2019 34th IEEE/ACM International Conference on Automated Software Engineering (ASE)}.\hskip 1em plus 0.5em minus 0.4em\relax IEEE/ACM, 2019, pp. 388--398.

\bibitem{alshahwan2024automated}
N.~Alshahwan, J.~Chheda, A.~Finogenova, B.~Gokkaya, M.~Harman, I.~Harper, A.~Marginean, S.~Sengupta, and E.~Wang, ``Automated unit test improvement using large language models at meta,'' in \emph{Companion Proceedings of the 32nd ACM International Conference on the Foundations of Software Engineering}, 2024, pp. 185--196.

\bibitem{xi2023rise}
Z.~Xi, W.~Chen, X.~Guo, W.~He, Y.~Ding, B.~Hong, M.~Zhang, J.~Wang, S.~Jin, E.~Zhou \emph{et~al.}, ``The rise and potential of large language model based agents: A survey,'' \emph{arXiv preprint arXiv:2309.07864}, 2023.

\bibitem{jimenez2024swebench}
\BIBentryALTinterwordspacing
C.~E. Jimenez, J.~Yang, A.~Wettig, S.~Yao, K.~Pei, O.~Press, and K.~R. Narasimhan, ``{SWE}-bench: Can language models resolve real-world github issues?'' in \emph{The Twelfth International Conference on Learning Representations}, 2024. [Online]. Available: \url{https://openreview.net/forum?id=VTF8yNQM66}
\BIBentrySTDinterwordspacing

\bibitem{mathews2024test}
N.~S. Mathews and M.~Nagappan, ``Test-driven development and llm-based code generation,'' in \emph{Proceedings of the 39th IEEE/ACM International Conference on Automated Software Engineering}, 2024, pp. 1583--1594.

\end{thebibliography}
\end{document}